\documentclass[%
reprint,
superscriptaddress,
bibnotes,
amsmath,amssymb,
aps,
prl,
longbibliography,
]{revtex4-2}

\usepackage{amssymb}
\usepackage{graphicx}
\usepackage{dcolumn}
\usepackage{bm}
\usepackage{chemformula}
\usepackage{hyperref}
\hypersetup{hidelinks}

\begin{document}


\title{Lattice symmetry breaking and strain control of single-$\mathbf{q}$ magnetism in \ch{Na_3Co_2SbO_6}}

\author{Yuchen~Gu}
\thanks{These authors contributed equally.}
\affiliation{International Center for Quantum Materials, School of Physics, Peking University, Beijing 100871, China}
\affiliation{Beijing National Laboratory for Condensed Matter Physics, Institute of Physics, Chinese Academy of Sciences, Beijing 100190, China}
\author{Wu~Xie}
\thanks{These authors contributed equally.}
\affiliation{Spallation Neutron Source Science Center, Dongguan 523803, China}
\affiliation{Institute of High Energy Physics, Chinese Academy of Sciences, Beijing 100049, China}
\author{Xintong~Li}
\affiliation{Beijing National Laboratory for Condensed Matter Physics, Institute of Physics, Chinese Academy of Sciences, Beijing 100190, China}
\author{Junhong~Liu}
\affiliation{International Center for Quantum Materials, School of Physics, Peking University, Beijing 100871, China}
\author{Xianghong~Jin}
\affiliation{International Center for Quantum Materials, School of Physics, Peking University, Beijing 100871, China}
\author{Dmitry~Khalyavin}
\affiliation{ISIS Facility, Rutherford Appleton Laboratory, Harwell Oxford, Didcot, OX11 0QX, United Kingdom}
\author{Fabio~Orlandi}
\affiliation{ISIS Facility, Rutherford Appleton Laboratory, Harwell Oxford, Didcot, OX11 0QX, United Kingdom}
\author{Pascal~Manuel}
\affiliation{ISIS Facility, Rutherford Appleton Laboratory, Harwell Oxford, Didcot, OX11 0QX, United Kingdom}
\author{Ping~Miao}
\email{miaoping@ihep.ac.cn}
\affiliation{Spallation Neutron Source Science Center, Dongguan 523803, China}
\affiliation{Institute of High Energy Physics, Chinese Academy of Sciences, Beijing 100049, China}
\author{Yuan~Li}
\email{yuan.li@iphy.ac.cn}
\affiliation{Beijing National Laboratory for Condensed Matter Physics, Institute of Physics, Chinese Academy of Sciences, Beijing 100190, China}
\affiliation{International Center for Quantum Materials, School of Physics, Peking University, Beijing 100871, China}
\date{October 20, 2026}
\begin{abstract}
Distinguishing single-$\mathbf{q}$ from multi-$\mathbf{q}$ magnetism constrains microscopic spin Hamiltonians. Here we determine the ground state of the candidate Kitaev magnet \ch{Na_3Co_2SbO_6} through its coupling to the lattice. High-resolution neutron diffraction reveals a spontaneous monoclinic-to-triclinic distortion at the magnetic transition, in contrast to the absence of detectable lattice-symmetry breaking in triple-$\mathbf{q}$ \ch{Na_2Co_2TeO_6}. Together with strain-selective magnetic responses, this identifies single-$\mathbf{q}$ order in \ch{Na_3Co_2SbO_6} over the proposed symmetry-preserving double-$\mathbf{q}$ state. Uniaxial compression along a low-symmetry in-plane direction achieves the domain selection that magnetic-field training could not, revealing the intrinsic field-induced phase boundaries of the dominant domain. These results establish strain as a symmetry-conjugate field for magnetic domains and highlight magnetoelastic coupling as a route to diagnose, manipulate, and potentially tune magnetic states.

\end{abstract}

\maketitle

Realizing a quantum spin liquid (QSL) \cite{BalentsNature2010,ZhouRMP2017,BroholmScience2020} based on the Kitaev model \cite{KitaevAP2006} requires not only sizable bond-dependent Ising interactions but also knowledge of the non-Kitaev terms that govern possible tuning routes toward the spin-liquid regime \cite{TakagiNRP2019,MotomeJPCM2020,TrebstPRRSP2022,MatsudaRMP2025}. High-spin $3d^7$ honeycomb cobaltates are attractive platforms in this regard because spin-orbit coupling produces low-energy pseudospin-$1/2$ moments, while competing nearest-neighbor superexchange channels can suppress Heisenberg exchange and retain strong bond-dependent interactions \cite{LiuPRB2018,SanoPRB2018,LiuPRL2020}. Experiments on \ch{Na_3Co_2SbO_6} (NCSO) and \ch{Na_2Co_2TeO_6} (NCTO) have provided evidence for substantial Kitaev-like interactions \cite{SongvilayPRB2020,SandersPRB2022}. NCTO, however, exposes an important complication: neutron spectroscopy and first-principles calculations point to substantial third-neighbor exchange \cite{YaoPRL2022,PandeyPRB2022}, while theoretical models show how six-spin ring exchange can select its triple-$\mathbf{q}$ ground state \cite{Kruger2023PRL,Wang2023PRB,Francini2024PRB}. These results motivate interactions beyond nearest-neighbor superexchange and raise the question of whether similarly complex Hamiltonians are needed across honeycomb cobaltates.

Central to this question is the proposal of multi-$\mathbf{q}$ magnetic order as a common ground-state motif in NCTO and NCSO. Specifically, the $C_3$-symmetric NCTO was assigned a triple-$\mathbf{q}$ ground state combining three symmetry-related zigzag components, while neutron diffraction on monoclinic NCSO suggested a double-$\mathbf{q}$ counterpart with two components \cite{ChenPRB2021,YaoPRR2023,Gu2024PRB}. These assignments, however, have different levels of experimental support. In NCTO, ordered-state excitations are inconsistent with conventional zigzag order and naturally captured by triple-$\mathbf{q}$ order \cite{ChenPRB2021,Kruger2023PRL}. Moreover, neither in-plane magnetic-field training nor $C_3$-breaking uniaxial strain repopulates putative zigzag domains \cite{YaoPRR2023,JinPRL2025}, and spin vorticity tied to a trainable ferrimagnetic moment provides a distinctive signature of the triple-$\mathbf{q}$ state \cite{YaoPRB2020,FranciniPRBferri2024,JinPRL2025}. In NCSO, by contrast, the double-$\mathbf{q}$ proposal rests mainly on two magnetic wave-vector components and their insensitivity to in-plane field training \cite{Gu2024PRB}. The null training result motivated this interpretation but does not exclude single-$\mathbf{q}$ domains that resist repopulation. An independent diagnostic comparable to the spin-vorticity response in NCTO is lacking, and monoclinic structural domains further complicate spectroscopy on large crystal arrays \cite{LiPRX2022}. A decisive symmetry-sensitive test is therefore needed, especially as the double-$\mathbf{q}$ assignment has already entered microscopic modeling \cite{BesthaPRB2026,mukharjeePRB2026}.

Magnetoelastic coupling offers such a test because the proposed multi-$\mathbf{q}$ states preserve the rotational symmetry of the paramagnetic lattice, whereas single-$\mathbf{q}$ order would lower the symmetry \cite{GuCPL2025}. This distinction has already motivated uniaxial-strain tests in NCTO \cite{JinPRL2025}, while spontaneous lattice-symmetry breaking (or its absence) at the magnetic ordering temperature $T_\mathrm{N}$ can be expected to provide a complementary zero-field probe. NCSO is particularly favorable for this approach: its in-plane magnetic anisotropy exceeds $200\%$ despite less than $0.2\%$ distortion of the honeycomb lattice \cite{LiPRX2022}, and pronounced anisotropic magnetostriction has been observed \cite{MiPRB2025,mukharjeePRB2026}. Resolving the lattice symmetry across $T_\mathrm{N}$ can therefore test the ordering wave-vector multiplicity, while applied strain might directly control the domain population if the order is single-$\mathbf{q}$.

\begin{figure}[!ht]
\includegraphics[width=3.375in]{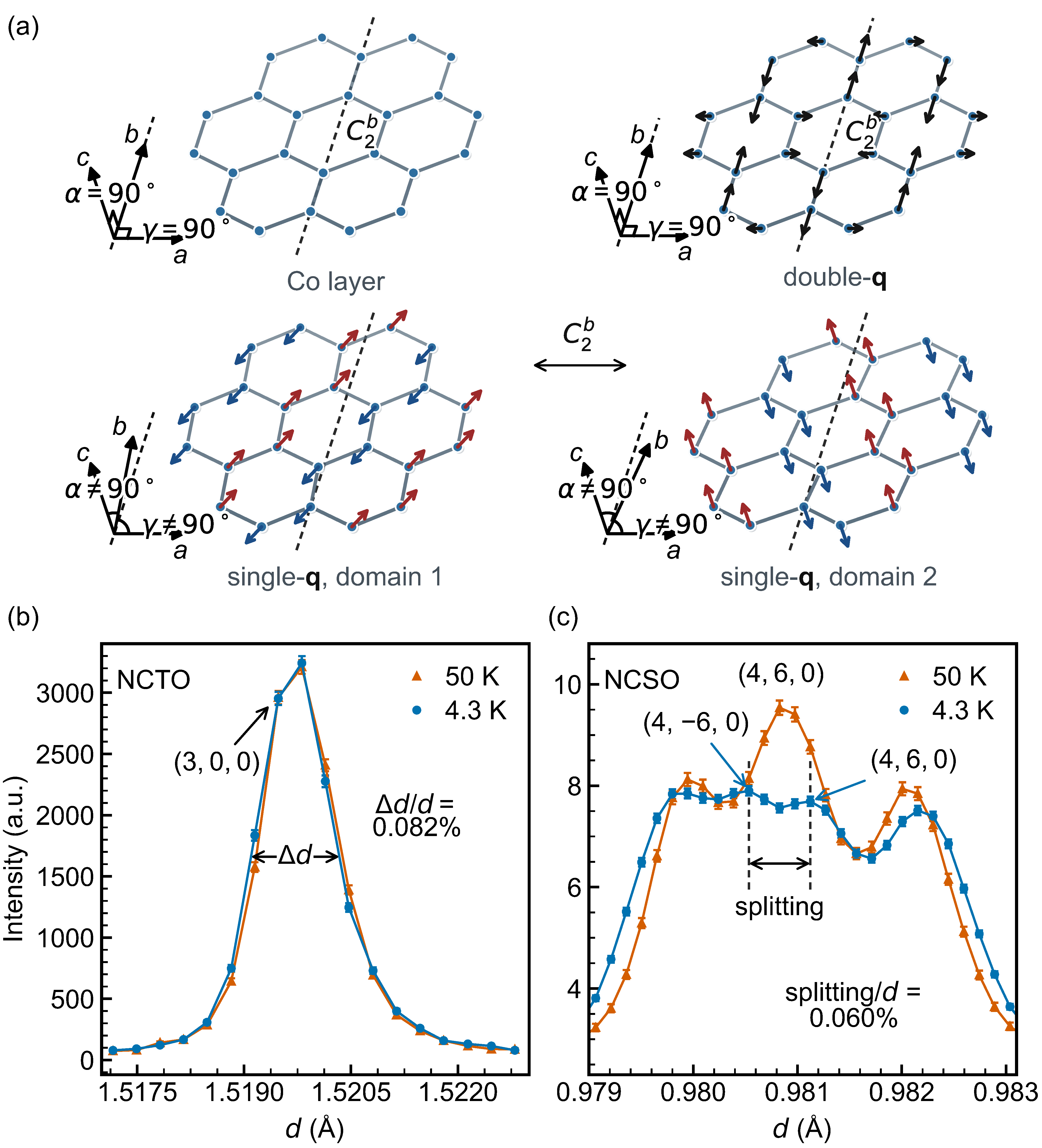}
\caption{Lattice-symmetry test of single-$\mathbf{q}$ and multi-$\mathbf{q}$ magnetism. (a) Schematics of the parent Co honeycomb layer in NCSO (upper left), the proposed symmetry-preserving double-$\mathbf{q}$ state (upper right), and the two $C_2^b$-related single-$\mathbf{q}$ domains (lower panels). Single-$\mathbf{q}$ order breaks $C_2^b$ and permits a triclinic distortion with $\alpha$ and/or $\gamma\ne90^\circ$. (b,c) High-resolution neutron diffraction profiles measured at TREND, CSNS \cite{TREND1,TREND2,TREND3}, at 50~K (orange triangles) and 4.3~K (blue circles): (b) the $(3,0,0)$ reflection of an NCTO single crystal and (c) the $(4,6,0)$ reflection of NCSO powder. NCTO shows no detectable splitting or additional broadening, whereas NCSO develops distinct $(4,6,0)$ and $(4,-6,0)$ contributions.}
\label{fig1}
\end{figure}

Here, we combine high-resolution neutron diffraction with strain-controlled magnetometry to distinguish single-$\mathbf{q}$ from double-$\mathbf{q}$ order in NCSO and resolve its intrinsic field response. The spontaneous triclinic distortion and strain-selectable domains identify a single-$\mathbf{q}$ ground state, removing ordering-vector multiplicity as a reason to invoke higher-order exchange. The contrast with unsuccessful field training illustrates how magnetoelastic coupling can expose magnetic order that conventional domain-selection methods leave unresolved.

Figure~\ref{fig1}(a) illustrates how lattice symmetry distinguishes the competing magnetic states in NCSO. Its distorted honeycomb lattice retains a twofold rotation about the $b$ axis, $C_2^b$, in place of the threefold symmetry of an ideal honeycomb layer \cite{ViciuJSSC2007}. Previous neutron diffraction identifies magnetic components at $(1/2,1/2,L)$ and $(1/2,-1/2,L)$, but not at $(0,1,L)$, in $C$-centered monoclinic notation \cite{Gu2024PRB,YanPRM2019}. The proposed equal-amplitude double-$\mathbf{q}$ superposition, with $\mathbf{q}=(1/2,\pm1/2,0)$, preserves the parent lattice point-group symmetry and therefore leaves symmetry-equivalent nuclear reflections degenerate in $d$ spacing. In contrast, single-$\mathbf{q}$ order selects one of the two $C_2^b$-related components, breaks this symmetry, and generically induces a triclinic lattice distortion through magnetoelastic coupling, with $\alpha$ and/or $\gamma$ departing from $90^\circ$. An analogous distinction applies to NCTO: single-$\mathbf{q}$ order lowers its $C_3^c$ lattice symmetry, whereas the triple-$\mathbf{q}$ state preserves it \cite{GuCPL2025}. Thus, spontaneous splitting of symmetry-equivalent nuclear reflections provides a test of the ordering-vector multiplicity.

High-resolution neutron diffraction can resolve the resulting differences in $d$ spacing, provided that suitable reflections are examined. For NCTO, any nonzero in-plane reflection can in principle reveal the loss of equivalence among its symmetry-related partners. For NCSO, sensitivity to the monoclinic-to-triclinic shear requires nonzero $K$ and at least one nonzero index among $H$ and $L$. For high-resolution neutron diffraction, NCSO single crystals were gently crushed and subsequently annealed to preserve crystalline quality and minimize sample-induced peak broadening without appreciably shifting $T_\mathrm{N}$ \cite{SM}. Figures~\ref{fig1}(b) and (c) compare representative profiles measured at 50 and 4.3~K, above and below the magnetic transitions of both compounds. NCTO shows no detectable splitting or additional broadening of its $(3,0,0)$ reflection upon cooling into the ordered state. In sharp contrast, NCSO exhibits a clear splitting of the $(4,6,0)$ reflection into the formerly degenerate $(4,6,0)$ and $(4,-6,0)$ contributions, with a relative separation of $6\times10^{-4}$. This contrast is consistent with symmetry-preserving triple-$\mathbf{q}$ order in NCTO and identifies broken $C_2^b$ symmetry in NCSO, supporting single-$\mathbf{q}$ order over the proposed $C_2^b$ symmetry-preserving double-$\mathbf{q}$ state.

\begin{figure}[!t]
\includegraphics[width=3.375in]{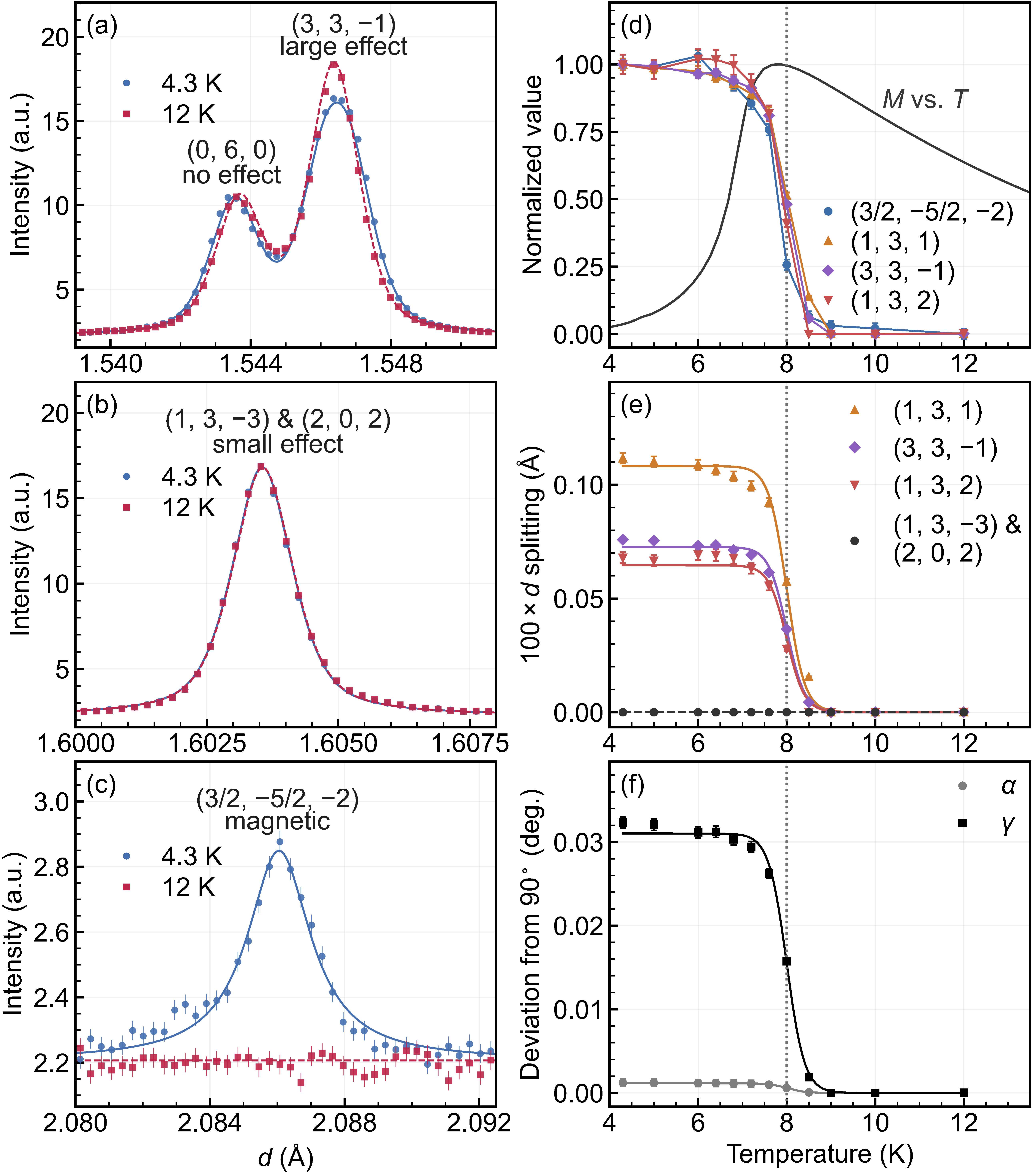}
\caption{Reflection dependence and magnetic origin of the triclinic distortion in NCSO. (a-c) Neutron powder diffraction profiles at 4.3 and 12~K: (a) the nuclear $(0,6,0)$ and $(3,3,-1)$ reflections, (b) the overlapping nuclear $(1,3,-3)$ and $(2,0,2)$ reflections, with comparable calculated intensities, and (c) the magnetic $(3/2,-5/2,-2)$ reflection. (d) Temperature dependence of the nuclear-peak splittings and magnetic Bragg intensity, together with the bulk magnetization, separately normalized. (e) Splittings of selected nuclear reflections and their description by the common triclinic lattice metric displayed in (f). Separate full-pattern Le Bail refinements yield $\alpha=90.0008(4)^\circ$ and $\gamma=90.0337(4)^\circ$ \cite{SM}. Vertical dashed lines in (d-f) mark $T_\mathrm{N}$.}
\label{fig2}
\end{figure}

Figure~\ref{fig2} presents the triclinic distortion's reflection-dependent signatures. The behavior varies substantially even among reflections with nonzero $H$ and $K$: $(3,3,-1)$ broadens appreciably, whereas no splitting is resolved for $(1,3,-3)$ within the measurement accuracy [Figs.~\ref{fig2}(a,b)]. This selectivity reflects the different sensitivities of the $d$ spacings to changes in $\alpha$ and $\gamma$. Fitting the reflection-dependent splittings yields the temperature evolution of $\alpha$ and $\gamma$ [Figs.~\ref{fig2}(e,f)]; independent full-pattern Le Bail refinements corroborate the triclinic distortion \cite{SM}. The same lattice metric accounts for the near-absence of splitting at $(1,3,-3)$. The departure of $\gamma$ from $90^\circ$ is much larger than that of $\alpha$, indicating that the symmetry-lowering distortion predominantly shears the honeycomb layers, with only a small change in their stacking geometry. This is consistent with the rotational-symmetry breaking within each layer caused by single-$\mathbf{q}$ zigzag order [Fig.~\ref{fig1}(a)].

The temperature dependence establishes the magnetic origin of this distortion. The splittings extracted from the nuclear peaks appear at $T_\mathrm{N}$ and, upon normalization, track the magnetic Bragg intensity within experimental uncertainty [Fig.~\ref{fig2}(d)]. Their onset also coincides with the antiferromagnetic transition seen in the bulk susceptibility. The lattice shear thus develops in proportion to the magnetic order parameter squared, demonstrating that it is driven by magnetic ordering. This coupling between zigzag order and in-plane shear suggests that uniaxial compression along a low-symmetry in-plane direction can favor one of the two magnetic domains, providing a route to strain-induced detwinning.

\begin{figure}[!t]
\includegraphics[width=3.375in]{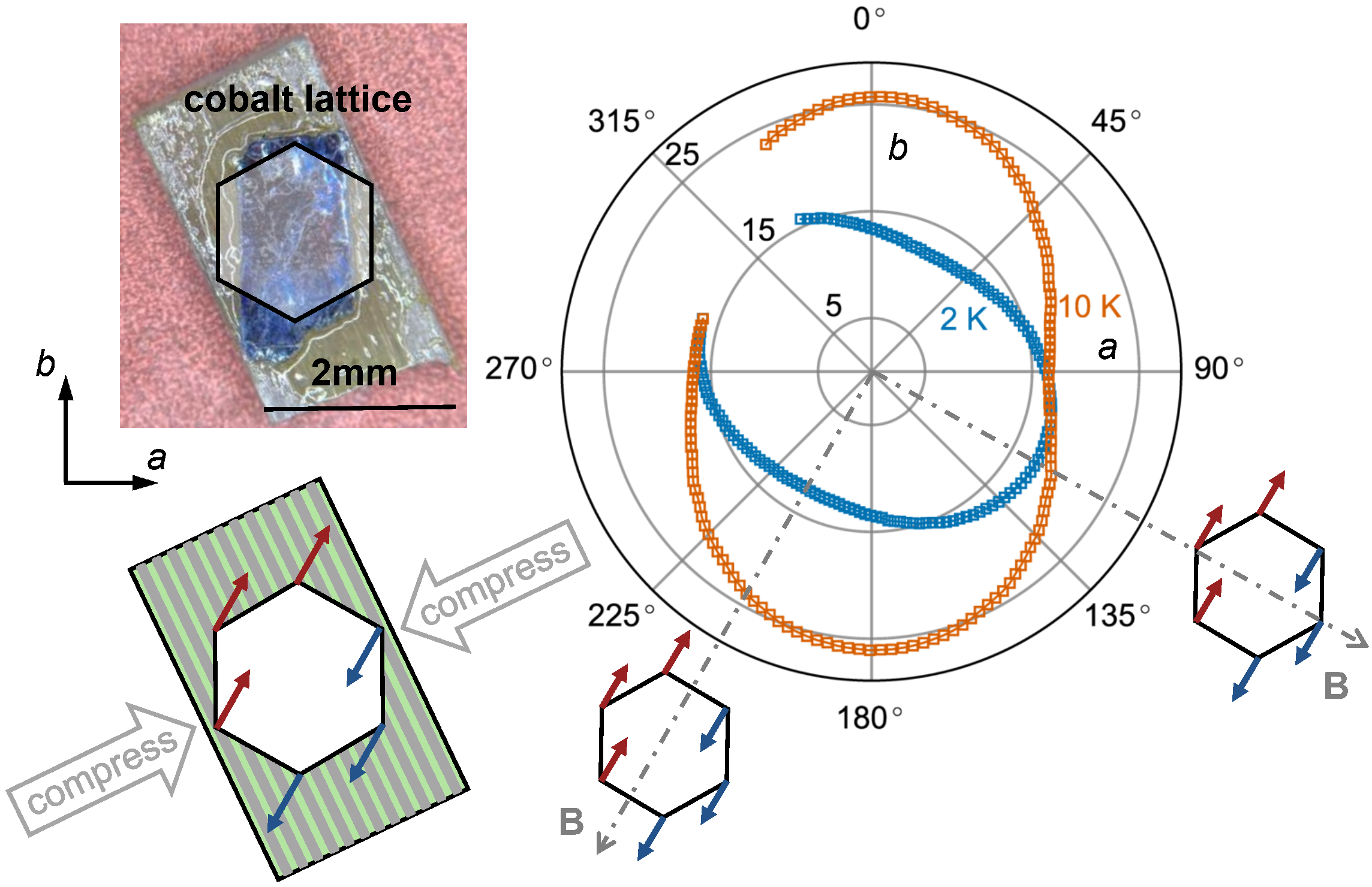}
\caption{Strain-induced selection of a zigzag domain in NCSO. Left: Photograph and schematic of the crystal bonded with GE varnish to a glass-fiber-reinforced plastic substrate. Gray arrows indicate the quasi-uniaxial compression generated perpendicular to the fibers upon cooling. Right: In-plane magnetic moment in units of $10^{-4}$~emu (radial scale), measured in a 0.1~T field at 10~K (orange) and 2~K (blue) as a function of angle from $b$. At fixed field, the moment is proportional to the susceptibility. At 2~K, the maximum response lies approximately $30^\circ$ from $a$, implying an ordered-moment direction approximately $30^\circ$ from $b$. The spin schematics illustrate fields parallel and perpendicular to the moments in the selected domain.}
\label{fig3}
\end{figure}

We test this possibility by bonding an NCSO crystal to a glass-fiber-reinforced plastic substrate \cite{HeNC2017}, whose anisotropic thermal contraction applies a small quasi-uniaxial compression along a low-symmetry in-plane direction upon cooling. Applied obliquely to the $a$ and $b$ axes, which have different compressibilities, this compression supplies a shear stress conjugate to the magnetically induced in-plane lattice shear. At 10~K, the susceptibility retains its maximum along $b$ and minimum along $a$, indicating that strain leaves the paramagnetic susceptibility axes essentially unchanged. At 2~K, by contrast, the maximum rotates to approximately $30^\circ$ from $a$ (Fig.~\ref{fig3}), a pronounced departure from the responses reported without deliberate strain \cite{LiPRX2022,BesthaPRB2026,mukharjeePRB2026} (strained NCTO retains a nearly isotropic in-plane magnetic response even in the ordered state \cite{SM}, as expected). The tilted response in NCSO is expected for a single zigzag domain: polarized neutron diffraction finds that each zigzag component has in-plane moments approximately $30^\circ$ from $b$ \cite{Gu2024PRB}, so a predominantly transverse susceptibility should peak perpendicular to them. Averaging equally populated $C_2^b$-related domains would instead restore susceptibility axes along $a$ and $b$. The observed tilted anisotropy axes indicate strong strain-induced detwinning, consistent with an estimated majority-to-minority domain ratio of about $16:1$ based on the angle-dependent magnetization results presented next \cite{SM}.

\begin{figure}[!t]
\includegraphics[width=3.375in]{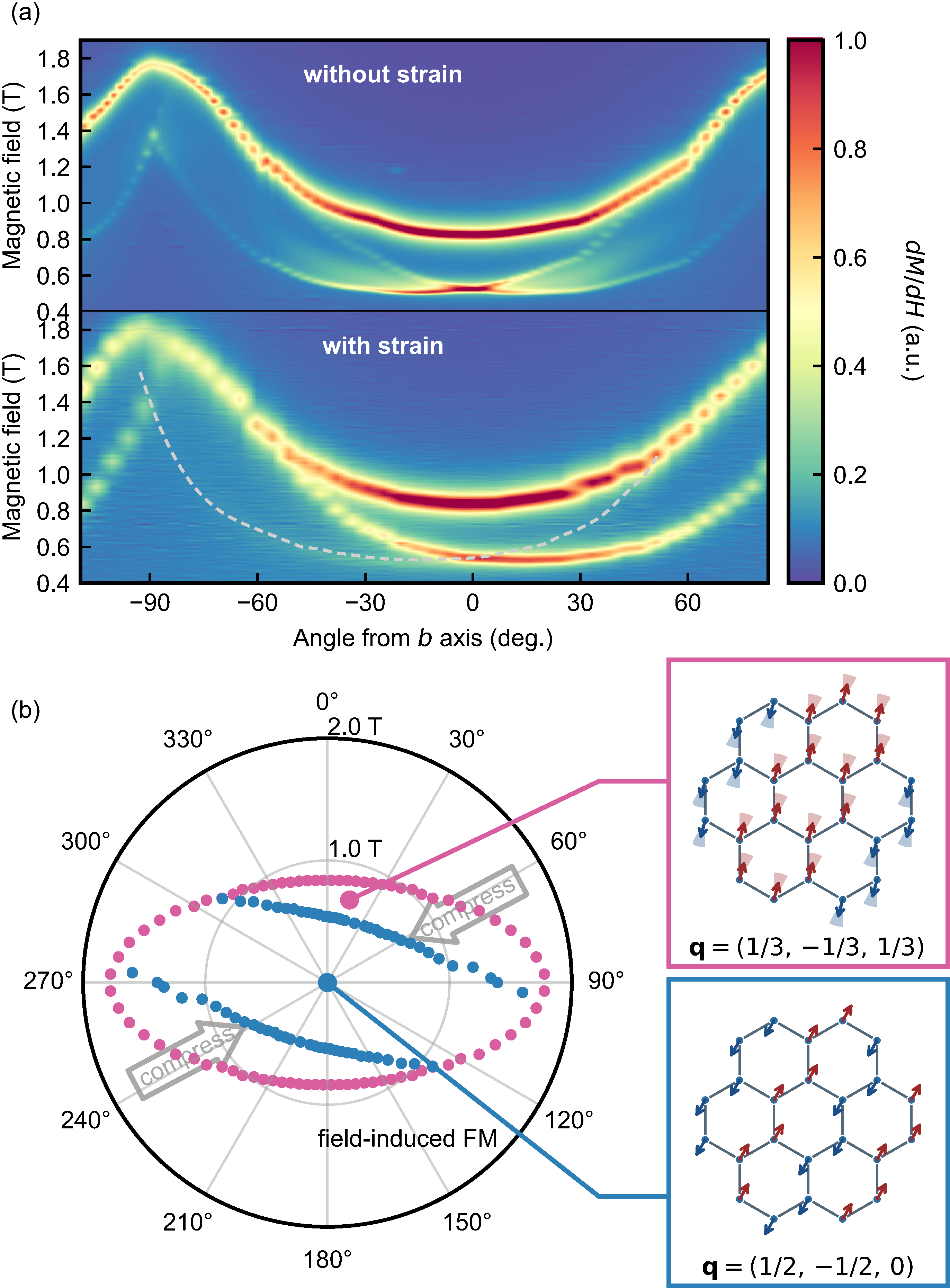}
\caption{Intrinsic field-induced phase boundaries revealed by detwinning. (a) Angular dependence of $dM/dH$ at 2~K with and without strain. The white dashed curve in the strained map marks the suppressed transition branch contributed by the other $C_2^b$-related domain. (b) Polar plot of transition fields extracted from the strained data: $\mathrm{AFM}_{1/2}$-$\mathrm{AFM}_{1/3}$ (blue) and $\mathrm{AFM}_{1/3}$-FM (pink), with angles measured from $b$. Insets show the zigzag $\mathrm{AFM}_{1/2}$ and up-up-down $\mathrm{AFM}_{1/3}$ arrangements for the selected domain; shaded sectors schematically illustrate flexible spin reorientation in $\mathrm{AFM}_{1/3}$.}
\label{fig4}
\end{figure}

Figure~\ref{fig4}(a) compares the angular dependence of $dM/dH$ at 2~K with and without strain, showing how detwinning resolves the multiple transition branches reported in unstrained NCSO \cite{LiPRX2022,Gu2024PRB,FanPRB2025} into the phase boundaries of the selected domain. The lower-field transition has two symmetry-related branches in the unstrained sample, whereas only one remains prominent after detwinning. The suppressed branch is thus the counterpart contributed by the other domain, rather than an additional transition of the selected domain. Strain therefore selects the domain population while leaving the measured transition fields essentially unchanged throughout the angular range studied.

The resulting phase boundaries [Fig.~\ref{fig4}(b)] distinguish the zigzag $\mathrm{AFM}_{1/2}$ phase, the intermediate $\mathrm{AFM}_{1/3}$ phase, and the field-induced FM phase. Following the neutron-diffraction assignments \cite{LiPRX2022,Gu2024PRB}, the two AFM phases have $\mathbf{q}=(1/2,-1/2,0)$ and $(1/3,-1/3,1/3)$ for the domain illustrated here. The near-$1/3$ magnetization plateau in $\mathrm{AFM}_{1/3}$ \cite{LiPRX2022} further supports the up-up-down pattern illustrated in Fig.~\ref{fig4}(b). The boundary between these two phases has a pronounced angular anisotropy tilted away from the $a$ and $b$ axes, as allowed by the broken $C_2^b$ symmetry of the selected zigzag domain. In contrast, the $\mathrm{AFM}_{1/3}$-FM boundary is nearly symmetric about these axes, despite its strong angular variation. WISH neutron diffraction at $15^\circ$ from $b$ likewise finds distinct lower but nearly coincident upper transition fields for the two domains \cite{SM}. This contrast is notable because our previous experimental evidence \cite{LiPRX2022, Gu2024PRB} indicates that the transition between the AFM phases preserves the in-plane propagation vector direction: the nearly symmetric upper boundary emerges even though $\mathrm{AFM}_{1/3}$ inherits a selected ordering direction. The two boundaries thus reveal distinct orientational constraints on the competing magnetic phases.

The single-$\mathbf{q}$ spin structure also constrains the exchange interactions. In the leading, bond-equivalent nearest-neighbor $J$-$K$-$\Gamma$-$\Gamma'$ model, \textit{i.e.}, treating the monoclinic structure only as a perturbation, the observed spin orientation along the in-plane-projected Kitaev axis of the bond connecting opposite spins requires $K_{\mathrm{eff}}=K+2\Gamma-2\Gamma'>0$ \cite{SM}. Neglecting the unresolved out-of-plane tilt \cite{Gu2024PRB} gives $K-\Gamma+\Gamma'\simeq0$ and hence $K_{\mathrm{eff}}\simeq3K>0$ within this model \cite{SM}, favoring antiferromagnetic Kitaev exchange. Monoclinicity lifts the equivalence of the three zigzag directions and can explain the absent domain with in-plane propagation along $b$. Bond counting clarifies the Kitaev interaction's energetic role: the collinear $\mathrm{AFM}_{1/2}$ and $\mathrm{AFM}_{1/3}$ patterns in Fig.~\ref{fig4} have, on average, one and two-thirds antiparallel nearest neighbors per site, respectively. With all three collinear patterns evaluated at the observed zero-field moment direction, the reference Kitaev energies per site are $E_K/S^2=-K/6$, $K/18$, and $K/2$ for $\mathrm{AFM}_{1/2}$, $\mathrm{AFM}_{1/3}$, and FM, respectively, where $S$ is the effective spin length. Positive $K$ thus favors zigzag over up-up-down order, and both over FM.

Antiferromagnetic further-neighbor Heisenberg exchange can lower both collinear AFM patterns relative to FM without selecting an in-plane spin direction \cite{SM}. At a common in-plane moment orientation, the up-up-down pattern retains two-thirds of the zigzag pattern's nearest-neighbor angular anisotropy. Its nearly $C_2^b$-symmetric boundary with FM therefore \textit{cannot be} explained by this exchange-energy comparison alone. The observed negligible off-axis tilt thus suggests spin reorientation as the field aligns the up-up-down pattern's net moment [Fig.~\ref{fig4}(b)]. This contrast hints at substantial isotropic exchange stabilizing $\mathrm{AFM}_{1/3}$ against FM and bond-dependent exchange fixing the zigzag spin direction. Intriguingly, the inferred orientational flexibility might make $\mathrm{AFM}_{1/3}$ more susceptible to quantum disorder than $\mathrm{AFM}_{1/2}$.

The symmetry of the domain imbalance clarifies why strain detwinning is effective. For zigzag amplitudes $\eta_1$ and $\eta_2$, the domain imbalance $Q=|\eta_1|^2-|\eta_2|^2$ and the in-plane shear $\varepsilon_{ab}$ are both even under time reversal and odd under $C_2^b$. A magnetic field is time-reversal odd, so the leading domain-selective free energy is $f_{\mathrm{sel}}=-\lambda\varepsilon_{ab}Q-\kappa H_aH_bQ$, with coupling coefficients $\lambda$ and $\kappa$. The strain term both accounts for the spontaneous shear and biases domains linearly, whereas the field term acts quadratically through the anisotropic susceptibility. In NCSO, the low transition fields limit the quadratic magnetic bias available before competing phases replace the zigzag state \cite{LiPRX2022,Gu2024PRB}. This restricted window can leave field training ineffective against domain pinning, for example from weak, spatially varying residual strain. Earlier field-training protocols failed in producing detectable remanent domain selection \cite{Gu2024PRB}, whereas measurements in an applied in-plane field reveal opposite intensity changes at satellites associated with the two zigzag propagation vectors \cite{SM}. Their implications are further discussed in the Supplemental Material \cite{SM}. An important advantage of strain is that it separates domain selection from magnetic-field tuning. This could provide a fixed domain reference for spectroscopy across field-driven transitions, helping distinguish changes in quantum dynamics from changes in domain population.

In summary, spontaneous triclinic distortion and strain-induced domain selection identify single-$\mathbf{q}$ zigzag order in NCSO over the proposed symmetry-preserving double-$\mathbf{q}$ state. Detwinning exposes phase boundaries with distinct angular symmetries, pointing to different orientational constraints in the zigzag and intermediate phases. By separating domain selection from magnetic-field tuning, strain provides a route to test microscopic models and explore quantum dynamics across field-induced phases with a controlled domain population.

\begin{acknowledgments}
We wish to thank Mingquan He, Giniyat Khaliullin, Huimei Liu, Je-Geun Park and John Tranquada for stimulating discussions. The work is supported in part by the National Key Research and Development Program of China (Grant Nos.~2025YFA1411500 and 2021YFA1401901) and by the National Natural Science Foundation of China (Grant Nos.~12474138 and 12304183). The work at the Spallation Neutron Source Science Center is further supported by the Guangdong Innovative \& Entrepreneurial Research Team Program (Grant No.~2021ZT09C539) and the Guangdong Basic and Applied Basic Research Foundation (Grant Nos.~2024B1515120045 and 2025B1515120034). The authors acknowledge the neutron beamtime at TREND (Shenzhen Guangming Science city Development and Construction Co.,Ltd. and its Material Genome Big-Science Facilities Platform) of the Chinese Spallation Neutron Source (CSNS, https://cstr.cn/31113.02.CSNS). The field-dependent neutron scattering experiment was performed at the ISIS facility in the United Kingdom under User Program RB2520127 \cite{WISHDATA}.
\end{acknowledgments}

\bibliography{NCSO_detwin}

\clearpage
\onecolumngrid
\setcounter{figure}{0}
\setcounter{table}{0}
\setcounter{equation}{0}
\setcounter{section}{0}
\renewcommand{\thefigure}{S\arabic{figure}}
\renewcommand{\thetable}{S\arabic{table}}
\renewcommand{\theequation}{S\arabic{equation}}
\renewcommand{\theHfigure}{supp.\arabic{figure}}
\renewcommand{\theHtable}{supp.\arabic{table}}
\renewcommand{\theHequation}{supp.\arabic{equation}}
\renewcommand{\theHsection}{supp.\arabic{section}}
\renewcommand{\theHsubsection}{supp.\arabic{section}.\arabic{subsection}}
\phantomsection
\label{sec:supplement}
\pdfbookmark[0]{Supplemental Material}{supplement}
\begin{center}
\textbf{\large Supplemental Material for ``Lattice symmetry breaking and strain control of single-$\mathbf{q}$ magnetism in \ch{Na_3Co_2SbO_6}''}
\end{center}

\section{Sample preparation}

\begin{figure}[htb]
\centering
\includegraphics[width=0.49\textwidth]{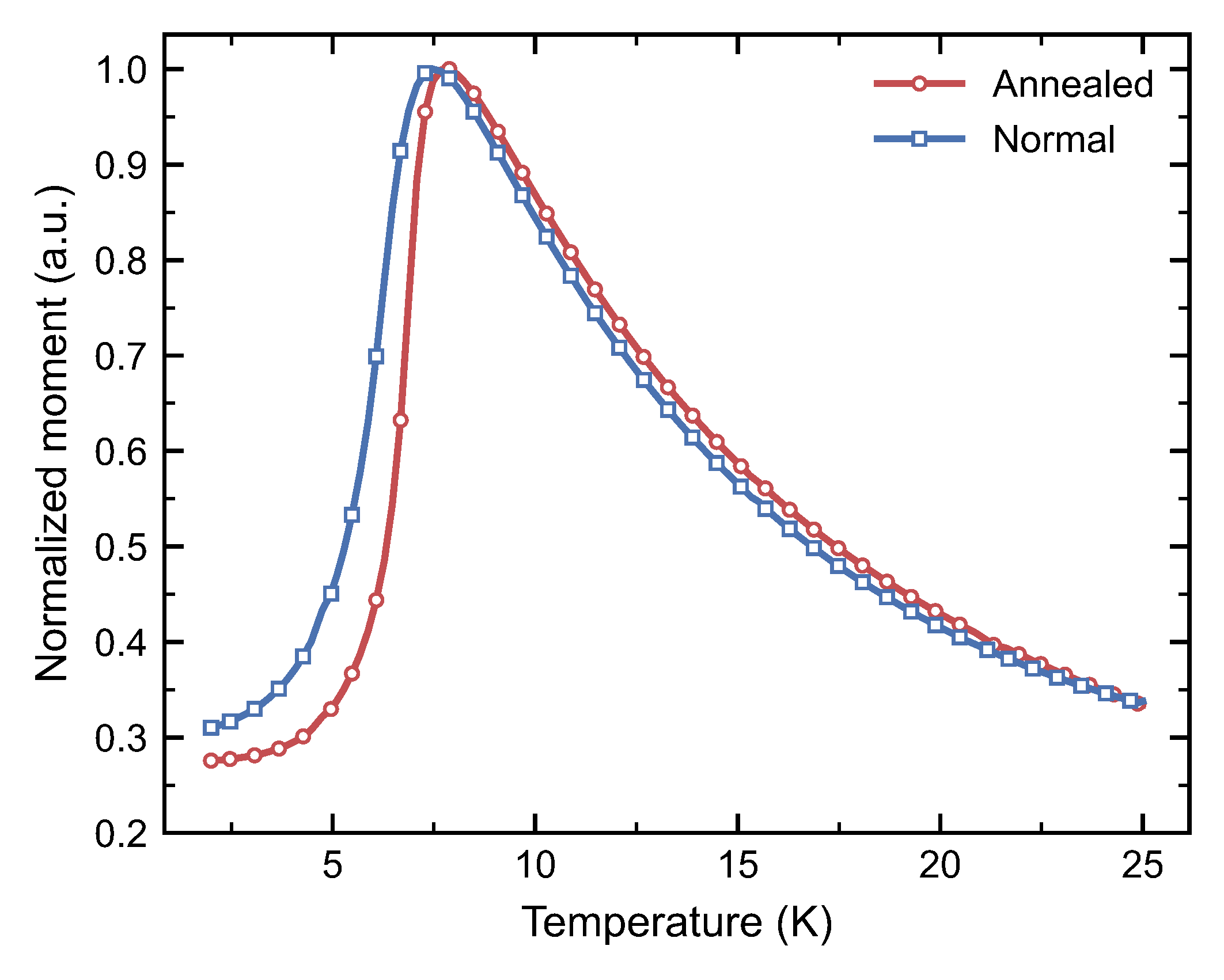}
\caption{Effect of annealing on NCSO powder. The magnetic moment measured in a field of 0.1~T is normalized to its maximum for each sample. Red circles denote annealed powder; blue squares denote the untreated powder labeled ``Normal'' in the plot. At this fixed field, the curves also represent the normalized magnetic susceptibility.}
\label{fig:anneal}
\end{figure}

The \ch{Na_2Co_2TeO_6} (NCTO) single crystal used for high-resolution neutron diffraction was from the batch studied in Ref.~\cite{JinPRL2025} and was annealed before measurement following the procedure described there to reduce residual strain. The \ch{Na_3Co2SbO_6} (NCSO) sample used for high-resolution neutron diffraction was prepared by gently crushing single crystals from the batch in Ref.~\cite{Gu2024PRB}, followed by annealing at $1000\,^\circ\mathrm{C}$ and cooling at $5\,^\circ\mathrm{C}/\mathrm{h}$. Fine grinding was avoided to preserve the crystalline quality of the fragments, while subsequent annealing was intended to reduce residual strain and defects introduced during fragmentation. This preparation protocol was chosen to minimize sample-induced peak broadening and facilitate the resolution of small symmetry-breaking peak splittings. The annealed and untreated samples have similar magnetic transition temperatures, while the annealed sample exhibits a sharper transition (Fig.~\ref{fig:anneal}), consistent with improved sample homogeneity.

The NCSO crystals used for strain-dependent magnetization and field-dependent neutron diffraction were also selected from the batch studied in Ref.~\cite{Gu2024PRB}. Polarization-resolved Raman measurements established the absence of monoclinic structural twinning, following Ref.~\cite{LiPRX2022}. Structural and magnetic domains should be distinguished: a single monoclinic structural domain can still contain both $C_2^b$-related zigzag domains. Strain selects between these zigzag domains in the measurements discussed in the main text.

\section{High-resolution neutron diffraction and lattice distortion}

\subsection{Measurements and full-pattern Le Bail refinements}

High-resolution neutron diffraction was performed on TREND at the China Spallation Neutron Source \cite{TREND1,TREND2,TREND3}. The best resolution in backscattering geometry reaches $\Delta d/d$ ($\Delta Q/Q$) $\simeq0.027\%$ at $2\theta = 177.3^\circ$. Sample-dependent contributions broaden the observed nuclear reflections: the effective relative peak widths (full width at half maximum versus $d$ divided by $d$) were approximately $0.08\%$ for the NCTO single crystal and $0.05\%$ for the NCSO powder. The latter allows the small NCSO peak splittings to be extracted by profile fitting.

Full neutron diffraction patterns of NCSO powder were collected at 4.35 and 50~K using a chopper frequency of 5~Hz over the neutron-wavelength range of 0--9.9~\AA{} (correspoding to approximately $d =$ 0--4.95~\AA{} at the backscattering detector bank), with approximately 2~h counting time per temperature. For the temperature-dependent measurements, the chopper frequency was increased to 25~Hz, yielding approximately five times higher flux over the neutron-wavelength range of 2.8--4.78~\AA{} (corresponding to approximately $d=1.4$--$2.4$~\AA{} at the backscattering detector bank). Patterns were collected at the base temperature of 4.35~K (labeled 4.3~K in the temperature-series plots), and at 5, 6, 6.4, 6.8, 7.2, 7.6, 8, 8.5, 9, 10, and 12~K, with 1.5~h counting time at each temperature.

\begin{figure}[p]
\centering
\includegraphics[width=0.82\textwidth]{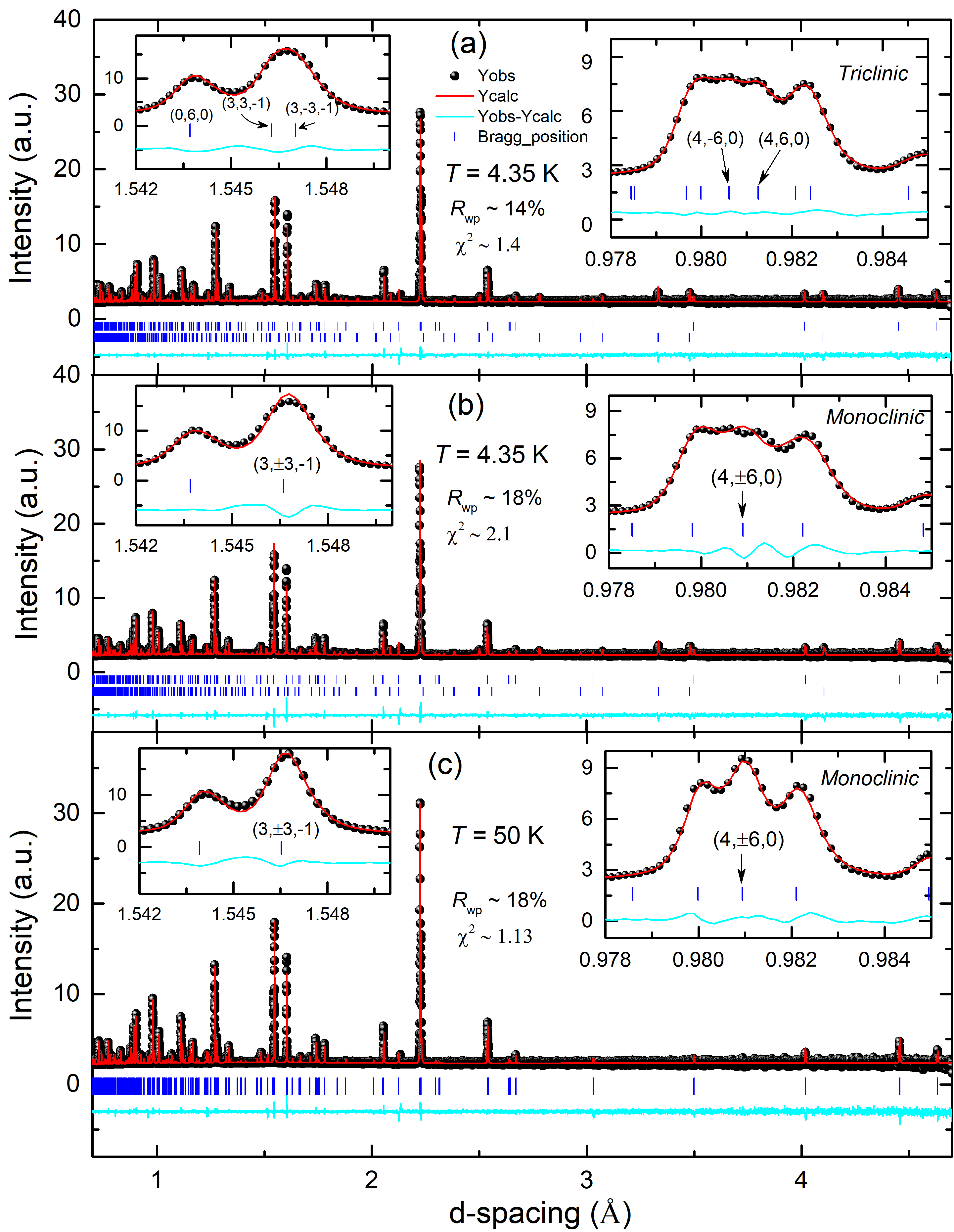}
\caption{Full-pattern Le Bail refinements of neutron diffraction from crushed NCSO single crystals. (a) Triclinic fit at 4.35~K, using the parent-cell setting described in the text. (b) Monoclinic $C2/m$ fit to the same 4.35~K data. (c) Monoclinic $C2/m$ fit at 50~K. Black points, red curves, and cyan curves denote observed intensities, calculated profiles, and their differences, respectively. The two rows of blue ticks mark nuclear and magnetic Bragg positions. The magnetic Bragg positions are described by the propagation vector $(1/2, 1/2, 0)$. Insets enlarge the $(3,\pm3,-1)$ and $(4,\pm6,0)$ regions; magnetic reflection markers are omitted from the insets because their intensities there are negligible. The triclinic metric resolves the low-temperature splitting of the nuclear reflections.}
\label{fig:lebail}
\end{figure}

The strong preferred orientation of the NCSO fragments motivated the use of Le Bail fitting rather than Rietveld refinement to determine the lattice metric and peak splittings without constraining the relative Bragg intensities through an atomic structural model. Full-pattern Le Bail fits provide an independent check of the triclinic distortion by describing all observed peaks in the measured range, complementing the analysis of individual peak splittings. Figure~\ref{fig:lebail} compares the triclinic and monoclinic fits to the same 4.35~K pattern, together with the monoclinic fit to the 50~K pattern. At 4.35~K, allowing a triclinic metric improves the description of the $(3,\pm3,-1)$ and $(4,\pm6,0)$ regions and reduces $R_{\mathrm{wp}}$ from approximately $18\%$ to $14\%$. The corresponding plotted goodness-of-fit values are $\chi^2\simeq2.1$ and $1.4$, respectively. At 50~K (above $T_\mathrm{N}$), the monoclinic fit describes the data with $R_{\mathrm{wp}}\simeq18\%$ and $\chi^2\simeq1.13$. The relatively large $R_{\mathrm{wp}}$ values shared by all fits reflect the limited counting statistics in the high-resolution detector bank, while the reflection-dependent residuals (cyan lines) provide a more direct comparison between the two low-temperature metrics. These results support a triclinic distortion on cooling. Corresponding insets compare the same $d$-spacing windows on the same plotting scale shared among the three panels of Fig.~\ref{fig:lebail}.

Table~\ref{tab:lattice} summarizes the fitted lattice parameters in a cell setting close to that of the parent monoclinic structure. This retains the parent indexing for comparison. The low-temperature deviations are $\alpha-90^\circ\simeq0.0008^\circ$ and $\gamma-90^\circ\simeq0.0337^\circ$, consistent with the distortion being dominated by in-plane shear.

\begin{table}[htb]
\caption{Lattice parameters from the Le Bail fits in Fig.~\ref{fig:lebail}. Parentheses give uncertainties in the last quoted digits. The monoclinic fits constrain $\alpha=\gamma=90^\circ$. All rows use the same parent-cell setting.}
\label{tab:lattice}
\begin{ruledtabular}
\begin{tabular}{lrrrrrr}
Model and temperature & $a$ (\AA) & $b$ (\AA) & $c$ (\AA) & $\alpha$ ($^\circ$) & $\beta$ ($^\circ$) & $\gamma$ ($^\circ$) \\
Triclinic, 4.35~K & 5.36053(2) & 9.26106(3) & 5.63282(2) & 90.0008(4) & 108.5736(2) & 90.0337(4) \\
Monoclinic, 4.35~K & 5.36038(2) & 9.26103(3) & 5.63285(2) & 90 & 108.5712(3) & 90 \\
Monoclinic, 50~K & 5.35973(2) & 9.26238(3) & 5.63311(2) & 90 & 108.5608(3) & 90
\end{tabular}
\end{ruledtabular}
\end{table}

\subsection{Temperature-dependent profile analysis}

The temperature dependence shown in Fig.~\ref{fig2} of the main text was obtained from joint fits of selected reflection groups, separately from the full-pattern Le Bail refinements. The groups near $d=1.545$, 1.740, and 2.227~\AA{} contain, respectively, $(3,3,-1)$ with $(0,6,0)$, $(1,3,2)$ with $(2,0,-3)$, and $(1,3,1)$ with $(2,0,-2)$. Each group contains a reflection sensitive to the triclinic shear and a neighboring reflection that does not split to first order in this distortion.

Each group was modeled with three pseudo-Voigt components: a doublet for the symmetry-related pair that splits below $T_{\mathrm N}$ and a single component for the neighboring reflection. All temperatures were fitted simultaneously. The doublet components had equal integrated intensities and positions symmetric about their common high-temperature position. Their intrinsic widths were shared across temperatures, while the width of the neighboring peak was allowed to vary. The total doublet intensity and the neighboring-peak intensity were each shared across temperatures. The three components within a group shared the same Lorentzian fraction. These constraints separate the small peak separation from changes in individual line-shape parameters. At each temperature, the fitted separations were described by a common triclinic metric, yielding the $\alpha$ and $\gamma$ values plotted in Fig.~\ref{fig2}(f) of the main text.

\subsection{Reflection-dependent sensitivity to the shear}

The sensitivity of a reflection can be obtained directly from the reciprocal metric. Let $\delta\alpha=\alpha-\pi/2$ and $\delta\gamma=\gamma-\pi/2$, with angular deviations expressed in radians. Linearizing about the monoclinic metric at fixed $a,b,c,\beta$ gives
\begin{equation}
\delta(d_{hkl}^{-2})=
\frac{2k}{b\sin^2\beta}
\left[
\left(\frac{h}{a}-\frac{l\cos\beta}{c}\right)\delta\gamma
+\left(\frac{l}{c}-\frac{h\cos\beta}{a}\right)\delta\alpha
\right].
\label{eq:metric}
\end{equation}
This term changes sign under $k\rightarrow-k$, lifting the monoclinic degeneracy of $(h,k,l)$ and $(h,-k,l)$. It vanishes when $k=0$, or when $h=l=0$, explaining the absence of first-order splitting for these reflections. Mixed indices permit a splitting but do not ensure that it will be resolvable, because the terms in Eq.~\eqref{eq:metric} can cancel.

For $(1,3,-3)$, the dominant $\delta\gamma$ coefficient is small because $h/a$ and $l\cos\beta/c$ nearly cancel. Because this cancellation suppresses the $\delta\gamma$ term, the $\delta\alpha$ term must also be retained despite the smaller angular distortion. Evaluating the full reciprocal metric using the triclinic parameters in Table~\ref{tab:lattice} gives relative separations of approximately $6.2\times10^{-6}$ for $(1,\pm3,-3)$ and $5.1\times10^{-4}$ for $(3,\pm3,-1)$, where $\Delta d$ is the separation of the two partners and $d$ their mean. Thus,
\begin{equation}
\frac{(\Delta d/d)_{(1,3,-3)}}{(\Delta d/d)_{(3,3,-1)}}\simeq0.012.
\label{eq:splitratio}
\end{equation}
The predicted $(1,3,-3)$ splitting is far smaller than the measured peak width. Its absence as a resolved doublet is therefore consistent with the same triclinic metric that describes the other reflections.

\section{Field-dependent neutron diffraction}
\subsection{Experimental conditions and reflection assignments}
\begin{figure}[!h]
\centering
\includegraphics[width=0.95\textwidth]{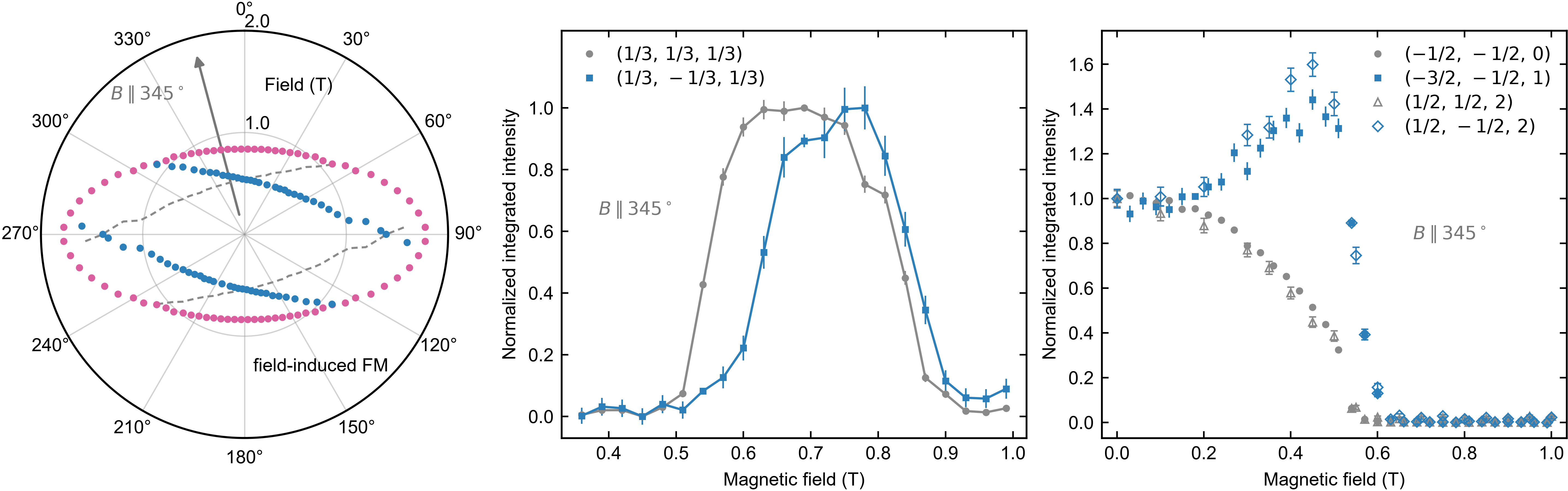}
\caption{Field-induced repopulation and domain-dependent transitions in NCSO. Left: Phase-boundary plot from Fig.~\ref{fig4}(b) of the main text, with the WISH field direction marked by a gray arrow. The angular origin is $b$, and the field is applied at $345^\circ~(-15^\circ)$. Middle: Field dependence of the integrated intensities of the intermediate-phase reflections at $(1/3,1/3,1/3)$ (gray circles) and $(1/3,-1/3,1/3)$ (blue squares), each normalized to its maximum. The onset fields differ, whereas the upper transition fields agree within the 0.03~T field step. Right: Integrated intensities of the zigzag magnetic satellites (-1/2,-1/2,0) and (1/2,1/2,2) from the $\mathbf q_1=(1/2,1/2,0)$ family (gray symbols), and (-3/2,-1/2,1) and (1/2,-1/2,2) from the $\mathbf q_2=(1/2,-1/2,0)$ family (blue symbols). Symbol shapes distinguish reflections as indicated in the legend. Each zigzag reflection is normalized to its own zero-field intensity. The disappearance of each half-integer family accompanies the appearance of the corresponding one-third family, rather than marking complete detwinning within the zigzag phase. All neutron data were measured at 1.5~K without deliberately applied strain. Error bars indicate one-standard-deviation statistical uncertainties.}
\label{fig:fieldneutrons}
\end{figure}

Field-dependent neutron diffraction was performed on WISH at the ISIS Neutron and Muon Source, UK \cite{WISH}. The NCSO crystal was measured without deliberately applied strain, with the magnetic field in the honeycomb plane at $-15^\circ$ from $b$, equivalent to $345^\circ$ in the angular convention of Fig.~\ref{fig4} of the main text and Fig.~\ref{fig:fieldneutrons}. This low-symmetry field direction also kept the target magnetic reflections within the vertical detector coverage of the cryomagnet geometry. Measurements were performed at 1.5 K, well below the magnetic ordering temperature, with counting times of 1.5 h at 0.21, 0.54 and 0.69 T and 8 min at all other fields. The longer counting times at 0.54 and 0.69 T account for the correspondingly smaller statistical uncertainties at these fields.

The middle panel of Figure~\ref{fig:fieldneutrons} compares reflections associated with the two ordering directions. The $(1/3,1/3,1/3)$ and $(1/3,-1/3,1/3)$ intensities develop at different fields on entering the intermediate phase, consistent with distinct lower phase boundaries for the two domains. Their high-field suppression, however, occurs over nearly the same field range: the inferred upper transition fields differ by less than the 0.03~T field step. This contrast supports the different angular symmetries of the lower and upper boundaries identified by magnetization in Fig.~\ref{fig4} of the main text.

The right panel of Figure~\ref{fig:fieldneutrons} compares four zigzag magnetic satellites. The reflections at (-1/2,-1/2,0) and (1/2,1/2,2) belong to the propagation-vector family $\mathbf q_1=(1/2,1/2,0)$, whereas those at (-3/2,-1/2,1) and (1/2,-1/2,2) belong to $\mathbf q_2=(1/2,-1/2,0)$. Each reflection is normalized to its own zero-field intensity. Throughout the following discussion, the two families are distinguished by their in-plane propagation directions; their out-of-plane propagation components change on entering the intermediate phase.

\subsection{Field-induced domain repopulation}

Below the transitions out of zigzag order, the $\mathbf q_1$ reflections decrease in intensity, whereas the $\mathbf q_2$ reflections increase with applying field. The similar overall field dependence of different satellites belonging to the same propagation-vector family supports a common change in the corresponding magnetic contribution, rather than a response peculiar to one reflection. Within the single-\textbf{q} description, this behavior is naturally interpreted as a field-induced increase of the $\mathbf q_2$ domain population at the expense of the $\mathbf q_1$ population.

The (1/2,-1/2,2) reflection provides a particularly useful test of an alternative explanation based on spin reorientation. For $\mathbf Q = (1/2,-1/2,2)$, the projection of $\mathbf Q$ onto the honeycomb plane is parallel to $\mathbf q_2$. Polarized neutron diffraction establishes that the zero-field staggered moments lie nearly entirely in the honeycomb plane and perpendicular to the corresponding in-plane propagation direction \cite{Gu2024PRB}. These moments are therefore perpendicular to the full $\mathbf Q$ in the planar limit, making the polarization factor essentially maximal. A substantial intensity enhancement at this reflection cannot be explained solely by rotating a fixed staggered moment to become more transverse to $\mathbf Q$.

The inferred repopulation is not complete before the field begins to replace zigzag order by the intermediate phase. Both half-integer reflection families remain measurable below the first conversion, and their subsequent disappearance must be considered together with the emergence of the one-third reflections. In particular, the loss of the $\mathbf q_1$ zigzag reflections is accompanied by the onset of (1/3,1/3,1/3) scattering. At a higher field, the $\mathbf q_2$ zigzag reflections disappear as the (1/3,-1/3,1/3) intensity develops. These changes correspond to the two propagation-vector conversions, which preserve the respective in-plane ordering directions while changing both the in-plane periodicity and the inter-layer modulation. The sequential zigzag-to-UUD transitions also argue against the scenario of unequal-amplitude double-\textbf{q} ground state. The sequence and field scales of these conversions are consistent with the two $H_{c1}$ branches in the unstrained magnetization at the same nominal field orientation [Fig.~\ref{fig4}(a) of the main text]. Within the single-\textbf{q} picture, these branches therefore correspond to domain-dependent zigzag-to-UUD transitions, rather than complete detwinning within the zigzag phase.

The evidence for repopulation within a single-\textbf{q} description should be distinguished from a model-independent determination of the zero-field ordering-vector multiplicity. An oblique in-plane field breaks the $C_2^b$ symmetry relating the two propagation-vector components and can favor unequal amplitudes via spin reorientation within a double-\textbf{q} state while both components remain nonzero. Such an amplitude change can produce opposite intensity variations and common scaling of several satellites belonging to the same propagation-vector family, without requiring domain repopulation.

Related contrasting responses of magnetic reflections have been reported for zigzag \ch{$\alpha$-RuCl_3} \cite{SearsPRB2017} and for NCTO \cite{YaoPRR2023}. In NCTO, the triple-\textbf{q} interpretation is supported by independent measurements \cite{JinPRL2025}. This comparison shows that contrasting field responses can occur in a multi-\textbf{q} material as well. Taken together with the lattice and strain results, the WISH data support the single-\textbf{q} domain interpretation and connect the two $H_{c1}$ branches to domain-dependent zigzag-to-UUD transitions. But the field-dependent intensities alone do not uniquely determine the zero-field ordering-vector multiplicity.

\newpage

\section{Strain-dependent magnetization and the NCTO comparison}

\subsection{NCSO domain selection}

For the NCSO magnetization measurements, a crystal was bonded with GE varnish to a glass-fiber-reinforced plastic substrate, following the thermal-contraction method of Ref.~\cite{HeNC2017}. Cooling generates quasi-uniaxial compression perpendicular to the fibers. The low-symmetry mounting orientation shown in Fig.~\ref{fig3} of the main text supplies a shear component in the crystallographic $ab$ plane and thus biases the two $C_2^b$-related zigzag domains. The low-field angular susceptibility was measured in 0.1~T at 10 and 2~K. At 2~K, the tilted susceptibility axes provide evidence for unequal domain populations. The suppression of one branch in the angle-dependent $dM/dH$ data of Fig.~\ref{fig4} provides complementary evidence for this imbalance.

\begin{figure}[!h]
\centering
\includegraphics[width=0.86\textwidth]{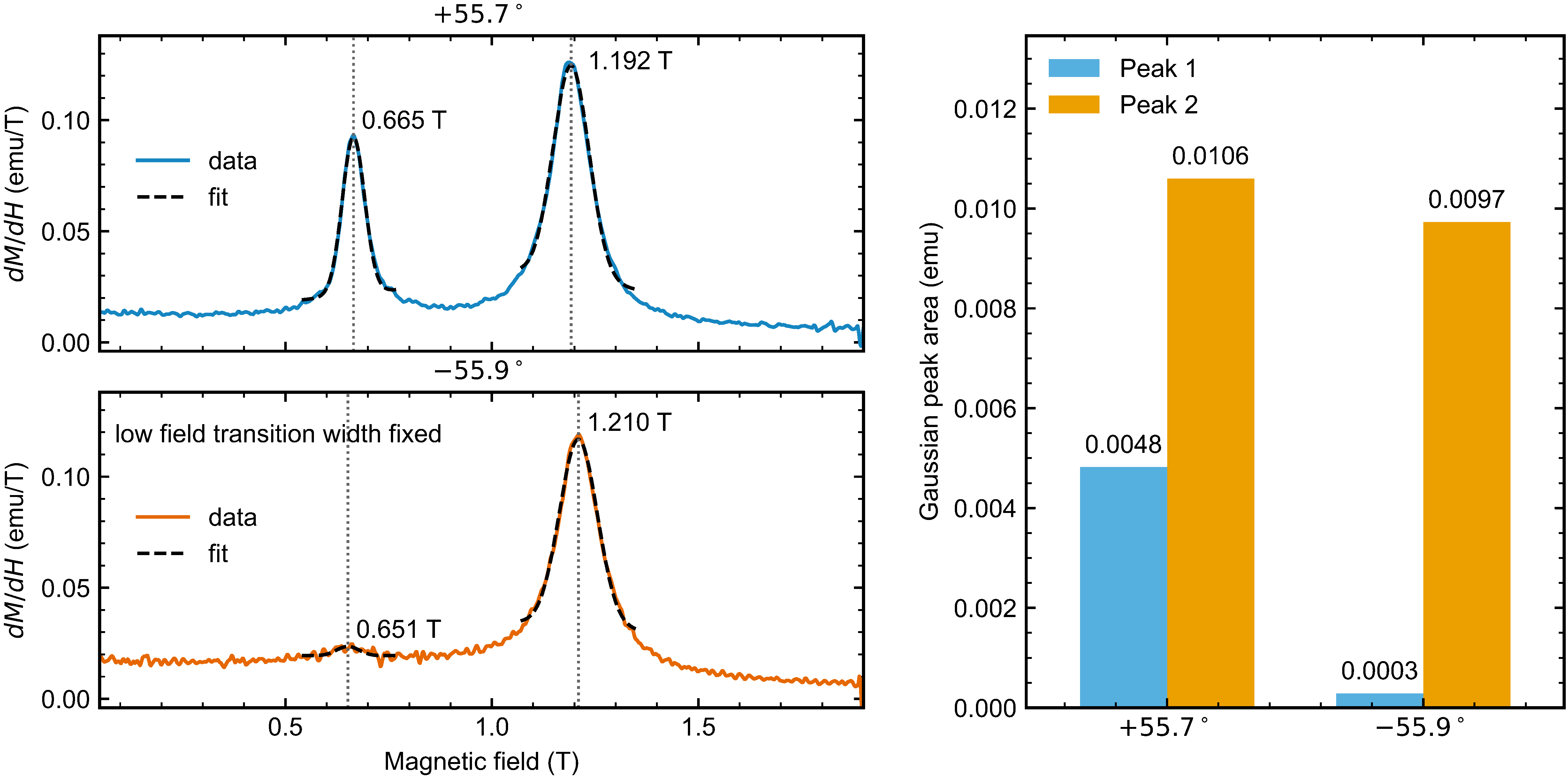}
\caption{Estimate of the magnetic-domain population of NCSO from the field-dependent magnetization. Left: $dM/dH$ measured at two approximately symmetry-related field orientations, $+55.7^\circ$ (top) and $-55.9^\circ$ (bottom), together with two-peak fits (black dashed curves). The vertical dotted lines mark the fitted transition fields. Because the low-field peak is strongly suppressed at $-55.9^\circ$, its width in the fit was fixed to that obtained at the symmetry-related $+55.7^\circ$ orientation. Right: Integrated Gaussian peak areas extracted from the fits. The low-field peak (Peak 1) decreases from 0.0048 to 0.0003 emu between the two orientations, corresponding to a peak-area ratio of approximately $16:1$, whereas the high-field peak (Peak 2) remains nearly unchanged (0.0106 and 0.0097 emu).}
\label{fig:NCSOdomainratio}
\end{figure}

A quantitative estimate of the domain imbalance can be obtained by comparing the transition amplitudes at two approximately symmetry-related field orientations, $+55.7^\circ$ and $-55.9^\circ$ from $b$, as displayed in Fig.~\ref{fig:NCSOdomainratio}. For equally populated $C_2^b$-related domains, the corresponding low-field transitions are expected to have comparable integrated peak areas. Instead, the integrated area of the low-field peak is reduced from 0.0048 emu at $+55.7^\circ$ to 0.0003 emu at $-55.9^\circ$, giving a ratio of approximately $16:1$. Interpreting this peak-area ratio as a population ratio assumes equal magnetization jumps per domain at the two symmetry-related orientations and negligible domain redistribution during measurement. In contrast, the high-field peak areas are nearly identical, 0.0106 and 0.0097 emu, respectively. The latter provides an internal check that the strong suppression of the low-field peak does not arise from an overall difference in signal magnitude between the two measurements.

\subsection{NCTO control measurements}

\begin{figure}[!h]
\centering
\includegraphics[width=0.85\textwidth]{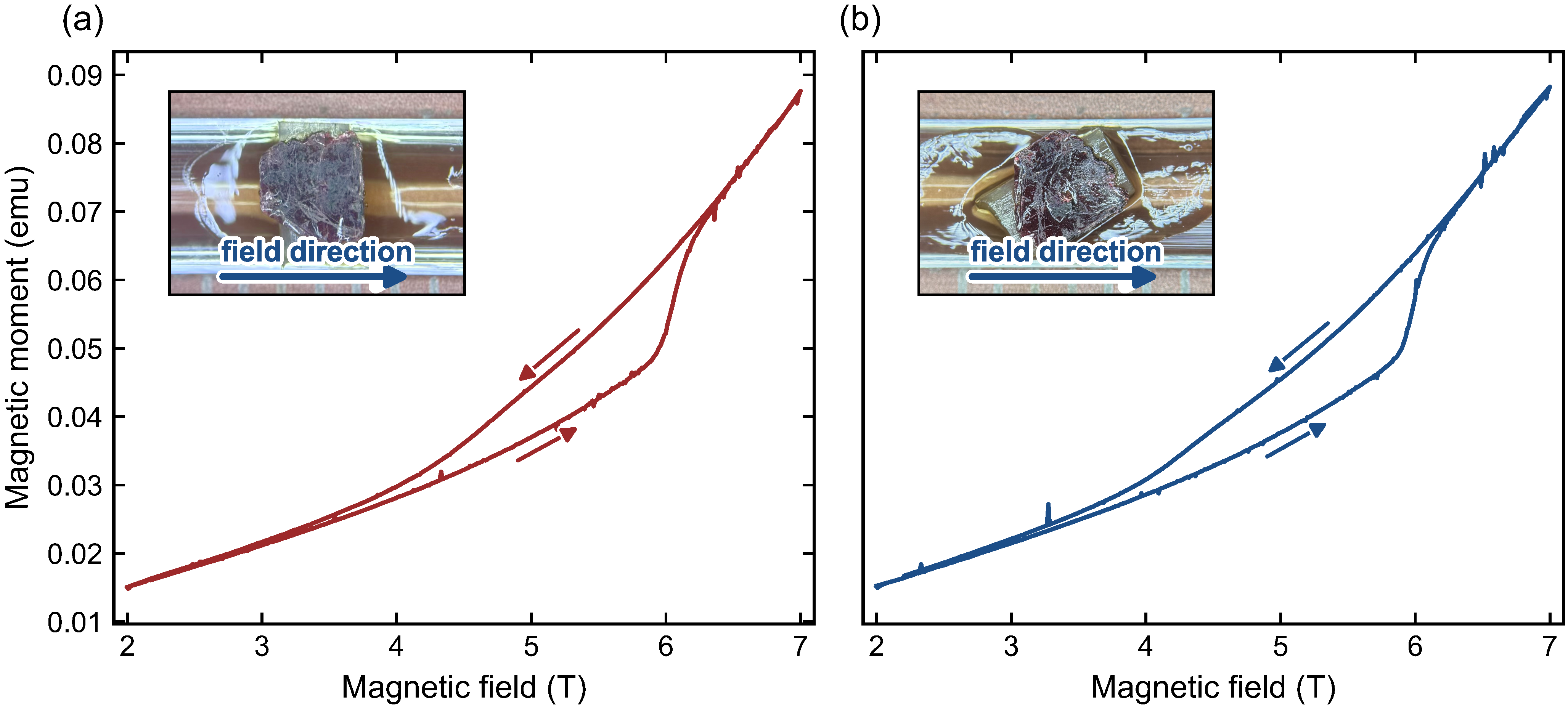}
\caption{Field-dependent magnetic moment of NCTO at 2~K for fields along two equivalent in-plane $a^*$ directions separated by $60^\circ$, shown in (a) and (b). The crystal and bonded glass-fiber substrate were rotated together between the two configurations. The insets show the mountings and field directions. The similar transition and hysteresis behavior indicates no appreciable strain-induced distinction between the two equivalent directions.
}
\label{fig:nctoM}
\end{figure}

\begin{figure}[!h]
\centering
\includegraphics[width=0.45\textwidth]{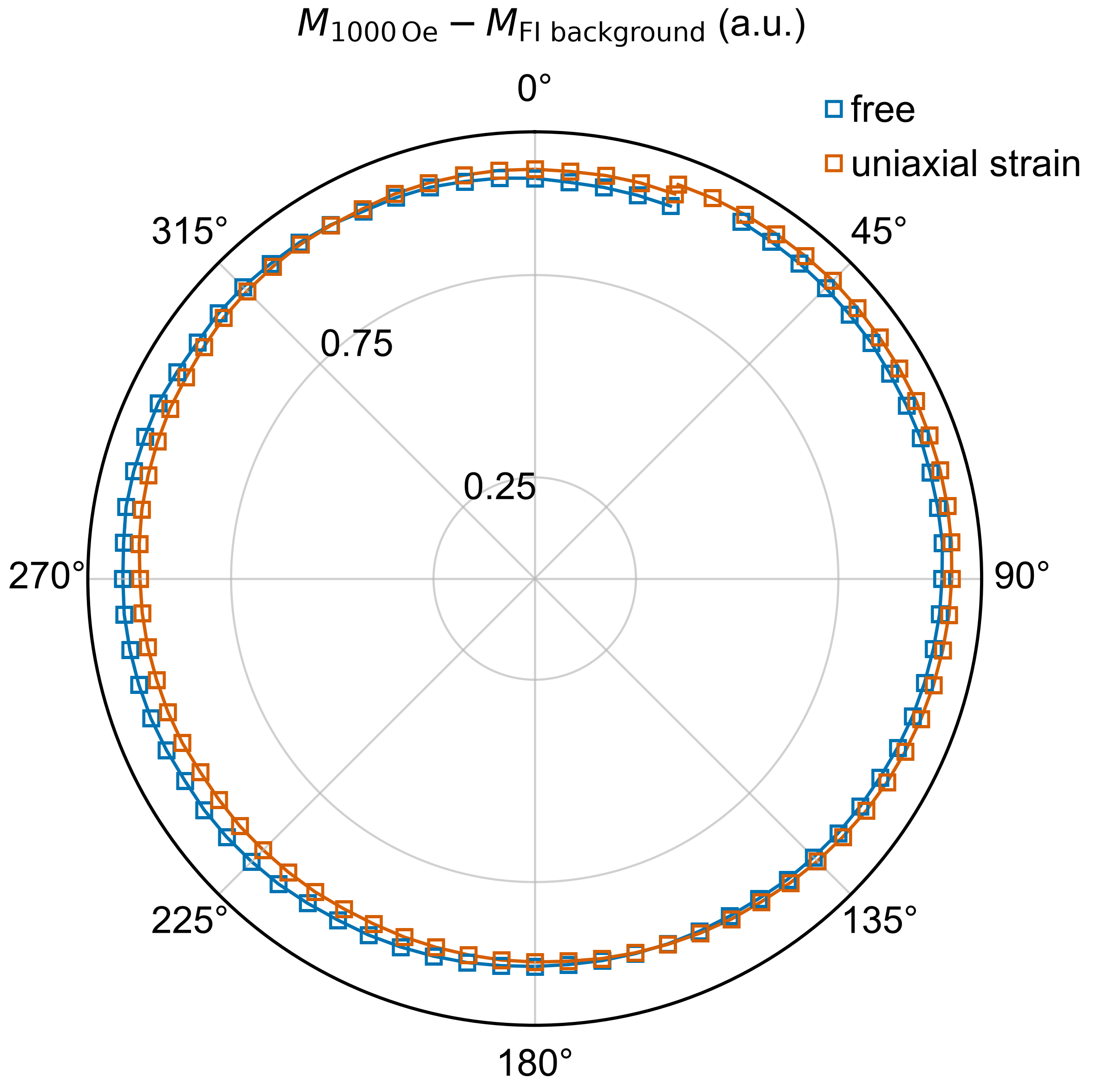}
\caption{In-plane angular magnetic response of NCTO at 2~K in a field of 1000~Oe (0.1~T), for a free crystal (blue) and under uniaxial strain (orange). The angular origin is $H\parallel a$. The ferrimagnetic background ($M_\mathrm{FI~background}$) was subtracted following the same protocol in Ref.~\cite{JinPRL2025}. The radial scale is in arbitrary units and, at fixed field, is proportional to the background-subtracted susceptibility. The free and strained curves are nearly identical. }
\label{fig:nctoangle}
\end{figure}

The NCTO measurements test whether the same approach produces the magnetic response expected from redistribution of putative single-$\mathbf q$ domains. Figure~\ref{fig:nctoM} compares the field-dependent magnetic moment along two crystallographically equivalent $a^*$ directions of hexagonal NCTO, separated by $60^\circ$. The crystal and its bonded glass-fiber substrate were rotated together, so the field changed direction relative to the imposed strain. The similar transition and hysteresis behavior in the two geometries shows that the substrate does not produce an appreciable distinction between these equivalent magnetic responses. The comparison is complemented by the low-field angular response at 2~K in Fig.~\ref{fig:nctoangle}: the free and strained measurements remain nearly coincident and show no pronounced new in-plane anisotropy. Here $0^\circ$ denotes $H\parallel a$. The ferrimagnetic background ($M_\mathrm{FI~background}$) was subtracted following the same protocol in Ref.~\cite{JinPRL2025}.

Together, these measurements reveal no appreciable strain-induced redistribution signature of the kind observed in NCSO. This null response is consistent with the symmetry-preserving triple-$\mathbf q$ state established for NCTO \cite{JinPRL2025}, which has no choice among the three single-$\mathbf q$ propagation directions for strain to select. This concerns orientational single-$\mathbf q$ domains; it does not exclude other domains, including those associated with the trainable spin vorticity of NCTO \cite{JinPRL2025}. The comparison complements the absence of detectable lattice-symmetry breaking in NCTO described in the main text.

\section{Constraints on the exchange model}

\subsection{Spin-direction selection in the bond-equivalent model}

We evaluate the spin-direction constraint within the leading bond-equivalent nearest-neighbor $J$--$K$--$\Gamma$--$\Gamma'$ model. This approximation neglects the differences among exchange tensors allowed by the monoclinic lattice. In local cubic spin coordinates, the Hamiltonian and the tensor on a $z$ bond are
\begin{equation}
\mathcal H_1=\sum_{\langle ij\rangle_\nu}\mathbf S_i^{\mathsf T}\mathcal J_\nu\mathbf S_j,
\qquad
\mathcal J_z=\begin{pmatrix}
J&\Gamma&\Gamma'\\
\Gamma&J&\Gamma'\\
\Gamma'&\Gamma'&J+K
\end{pmatrix},
\label{eq:H1}
\end{equation}
with $\nu=x,y,z$ and cyclic permutations defining $\mathcal J_x$ and $\mathcal J_y$ \cite{LiuPRB2018}. Each bond is counted once, and $K>0$ denotes antiferromagnetic Kitaev exchange.

\begin{figure}[htb]
\centering
\includegraphics[width=0.8\textwidth]{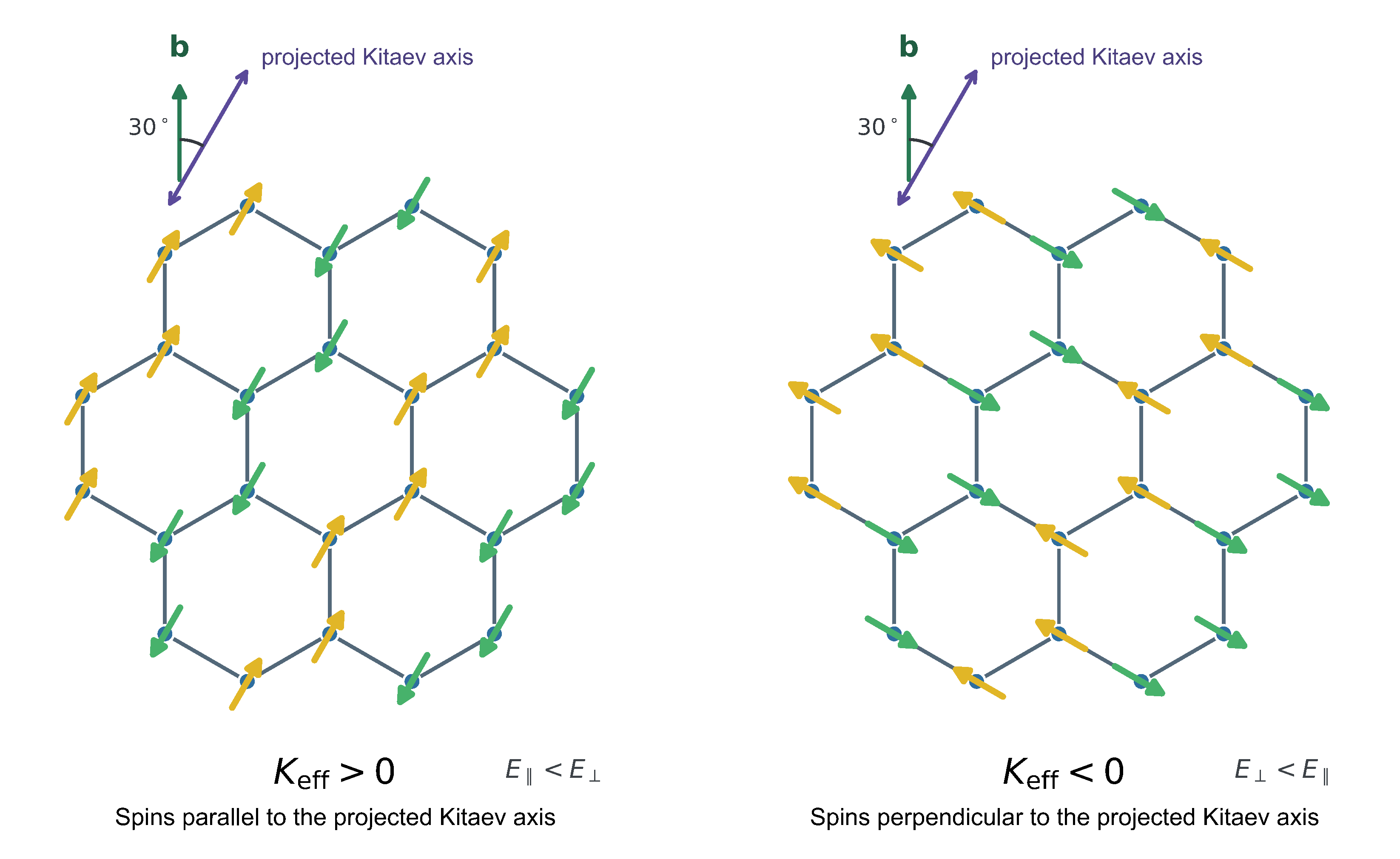}
\caption{Two principal in-plane moment directions for the same zigzag pattern. Left: Spins parallel to the projected Kitaev axis, favored for $K_{\mathrm{eff}}>0$. Right: The orthogonal orientation, favored for $K_{\mathrm{eff}}<0$. The projected axis lies approximately $30^\circ$ from $b$ for the domain shown. The sign criterion applies within the bond-equivalent model defined in Eq.~\eqref{eq:H1}.}
\label{fig:spins}
\end{figure}

Consider a zigzag domain with ferromagnetic correlations on the $x$ and $y$ bonds and antiferromagnetic correlations on the $z$ bonds. For $\mathbf S_i=\sigma_i S\hat{\mathbf n}$, with $\sigma_i=\pm1$, its classical energy per site is
\begin{equation}
E_{1/2}(\hat{\mathbf n})=\frac{S^2}{2}\hat{\mathbf n}^{\mathsf T}\mathcal A_z\hat{\mathbf n},
\qquad \mathcal A_z=\mathcal J_x+\mathcal J_y-\mathcal J_z.
\label{eq:A}
\end{equation}
The ideal layer normal and two orthogonal in-plane directions are
\begin{equation}
\hat{\mathbf e}_n=\frac{(1,1,1)}{\sqrt3},\qquad
\hat{\mathbf e}_\parallel=\frac{(-1,-1,2)}{\sqrt6},\qquad
\hat{\mathbf e}_\perp=\frac{(1,-1,0)}{\sqrt2}.
\label{eq:axes}
\end{equation}
Here $\hat{\mathbf e}_n$ denotes the layer normal, not the monoclinic direct-lattice $c$ axis. The vector $\hat{\mathbf e}_\parallel$ is the in-plane projection of the local $z$ Kitaev axis. For the selected zigzag domain, it corresponds to the observed moment direction approximately $30^\circ$ from crystallographic $b$ (Fig.~\ref{fig:spins}).

In the basis $\{\hat{\mathbf e}_\perp,\hat{\mathbf e}_\parallel,\hat{\mathbf e}_n\}$, the exchange matrix is
\begin{equation}
\widetilde{\mathcal A}_z=
\begin{pmatrix}
A_\perp&0&0\\
0&A_\parallel&B\\
0&B&A_n
\end{pmatrix},
\label{eq:rotated}
\end{equation}
where
\begin{align}
A_\perp&=J+K+\Gamma-2\Gamma', &
A_\parallel&=J-\frac K3-\frac{5\Gamma}3+\frac{2\Gamma'}3,\nonumber\\
A_n&=J+\frac K3+\frac{2\Gamma}3+\frac{4\Gamma'}3, &
B&=\frac{2\sqrt2}{3}(\Gamma-\Gamma'-K).
\label{eq:elements}
\end{align}
Writing an in-plane direction as $\hat{\mathbf n}(\phi)=\cos\phi\,\hat{\mathbf e}_\parallel+\sin\phi\,\hat{\mathbf e}_\perp$ gives
\begin{equation}
E_{1/2}(\phi)=E_\parallel+\frac{2S^2}{3}K_{\mathrm{eff}}\sin^2\phi,
\qquad K_{\mathrm{eff}}=K+2\Gamma-2\Gamma'.
\label{eq:keff}
\end{equation}
Thus the observed in-plane direction selects $K_{\mathrm{eff}}>0$ in this approximation.

A rotation out of the plane, $\hat{\mathbf n}(\theta)=\cos\theta\,\hat{\mathbf e}_\parallel+\sin\theta\,\hat{\mathbf e}_n$, has energy
\begin{equation}
E_{1/2}(\theta)=\frac{S^2}{2}\left(A_\parallel\cos^2\theta+A_n\sin^2\theta+2B\sin\theta\cos\theta\right).
\label{eq:tilt}
\end{equation}
An exactly in-plane stationary moment requires $B=0$, or $K-\Gamma+\Gamma'=0$. Combining this relation with Eq.~\eqref{eq:keff} yields $K_{\mathrm{eff}}=3K>0$. Stability against an out-of-plane rotation additionally requires $A_n>A_\parallel$. These are conditions on the orientation of the chosen zigzag pattern, not a proof that it is the global ground state.

For a small unresolved tilt, the stationary condition is
\begin{equation}
\tan2\theta=\frac{2B}{A_\parallel-A_n}.
\label{eq:smalltilt}
\end{equation}
Neglecting this tilt gives the approximate relation $K_{\mathrm{eff}}\simeq3K>0$ used in the main text. A quantitative tolerance on the coupling relation requires a bound on $\theta$ and knowledge of the out-of-plane stiffness. Monoclinic bond inequivalence, additional anisotropic interactions, and fluctuation corrections can modify the inference of antiferromagnetic Kitaev exchange.

\subsection{Reference energies of the collinear field-induced patterns}

To compare the patterns illustrated in Fig.~\ref{fig4} of the main text, we take ferromagnetic zigzag chains with alternating signs in $\mathrm{AFM}_{1/2}$, repeating up-up-down signs in $\mathrm{AFM}_{1/3}$, and uniform signs in FM. The nearest-neighbor bond correlations $(c_x,c_y,c_z)=\langle\sigma_i\sigma_j\rangle$ are then
\begin{equation}
(c_x,c_y,c_z)_{1/2}=(1,1,-1),\qquad
(c_x,c_y,c_z)_{1/3}=(1,1,-1/3),\qquad
(c_x,c_y,c_z)_{\mathrm{FM}}=(1,1,1).
\label{eq:correlations}
\end{equation}
The Kitaev energy per site is $E_K=(KS^2/2)\sum_\nu c_\nu n_\nu^2$. Evaluating all three patterns at the observed zero-field direction $\hat{\mathbf e}_\parallel$ gives
\begin{equation}
\frac{1}{S^2}(E_{K,1/2},E_{K,1/3},E_{K,\mathrm{FM}})=\left(-\frac K6,\frac K{18},\frac K2\right).
\label{eq:kitaevenergies}
\end{equation}
These are reference exchange energies at a common orientation, rather than the relaxed energies of the phases under an applied field.

For isotropic first-, second-, and third-neighbor exchanges $J$, $J_2$, and $J_3$, counting each pair once gives
\begin{align}
E_{\mathrm{iso},1/2}/S^2&=J/2-J_2-3J_3/2,\nonumber\\
E_{\mathrm{iso},1/3}/S^2&=5J/6+J_2/3-J_3/2,\nonumber\\
E_{\mathrm{iso},\mathrm{FM}}/S^2&=3J/2+3J_2+3J_3/2.
\label{eq:isotropic}
\end{align}
For antiferromagnetic $J_2,J_3>0$, the further-neighbor contribution therefore lowers the energies per site relative to FM by $S^2(4J_2+3J_3)$ for zigzag and $S^2(8J_2/3+2J_3)$ for the up-up-down pattern. These gains are independent of the common spin direction and underlie the energetic tendency discussed in the main text. The consequences for the stability of the intermediate phase require a separate comparison, as discussed below.

For the bond-equivalent nearest-neighbor tensors, the up-up-down energy at any common spin direction obeys
\begin{equation}
E_{1/3}(\hat{\mathbf n})=\frac23E_{1/2}(\hat{\mathbf n})+\frac13E_{\mathrm{FM}}(\hat{\mathbf n}).
\label{eq:interpolation}
\end{equation}
The FM exchange energy is independent of the in-plane direction in this model. Consequently, the up-up-down pattern retains two-thirds of the zigzag in-plane angular anisotropy. Its nearly symmetric upper phase boundary cannot be attributed to cancellation of this anisotropy at fixed spin direction, motivating the spin-reorientation interpretation in the main text.

Equation~\eqref{eq:interpolation} also holds for the isotropic energies in Eq.~\eqref{eq:isotropic}. The rigid collinear patterns have mean spins $0$, $S/3$, and $S$ along their common axis, so the same interpolation holds after adding the Zeeman energy. Bond counting of these trial patterns alone therefore does not produce a finite field interval in which the up-up-down state lies below both competitors. A quantitative account of its stability requires physics beyond this comparison, such as noncollinear relaxation, interlayer correlations, fluctuation effects, or additional interactions. The energy comparisons used here constrain microscopic tendencies without constituting a complete calculation of the field-induced phase diagram.

\end{document}